# Atomic-Scale Insights into Solute Drag Effects on Grain Boundary Motion in Mg-Al and Mg-Ca Alloys

Zhishun Chen[1], Shudong He[1], Shuai Zhang[1], Xiaohan Bie[2], Zhuoming Xie[3], Tengfei Yang[1], Wangyu Hu[1], Huiqiu Deng[4], Shiwei Xu[5*], Zhuoran Zeng[1*], Jie Hou[1*]

1. College of Materials Science and Engineering, State Key Laboratory of Cemented Carbide, Hunan University, Changsha 410082, China
2. Department of Mining and Materials Engineering, McGill University, Montreal, Quebec, H3A 0C5, Canada
3. Key Laboratory of Materials Physics, Institute of Solid State Physics, Chinese Academy of Sciences, P. O. Box 1129, Hefei 230031, PR China
4. School of Physics and Electronics, Hunan University, Changsha 410082, China
5. State Key Laboratory of Advanced Design and Manufacturing Technology for Vehicle, Hunan University, Changsha, 410082, China

* Corresponding authors: xushiwei@hnu.edu.cn (Shiwei Xu), zeng.zhuoran@hnu.edu.cn (Zhuoran Zeng), jiehou@hnu.edu.cn (Jie Hou)

**Abstract**

The slip behavior of dislocations and grain boundaries critically governs recrystallization and plastic deformation in Mg alloys and can be strongly influenced by solutes. However, the quantitative effects of solute distribution on defect mobility remain unclear. Using molecular dynamics and Monte Carlo simulations, we systematically investigate how Al and Ca solutes affect the motion of dislocations, low-angle grain boundaries (LAGBs), and high-angle grain boundaries (HAGBs) in Mg. Within the idealized framework of random solid-solution, solute drag is dominated by elastic interactions arising from atomic size mismatch, resulting in a stronger resistance from Ca than from Al. In contrast, under the more realistic condition where solute segregation occurs, the dominant mechanism shifts to chemically driven pinning, whose effectiveness is governed by the attainable segregation density. Owing to strong Ca-Ca repulsion, Al achieves substantially higher segregation concentrations than Ca and therefore exerts much stronger pinning effects. Notably, solute-induced retardation is significantly more pronounced for HAGBs than for LAGBs, leading to amplified solute effects during the late stages of recrystallization, where grain growth is controlled primarily by HAGB migration. These results provide atomic-scale insight into experimentally observed grain refinement in Mg alloys.

# 1. Introduction

Magnesium (Mg) alloys, distinguished as the lightest structural metallic materials, possess high specific strength, superior stiffness, and excellent recyclability, rendering them critical candidates for weight reduction in transportation, aerospace, and electronics sectors [1-4]. However, the intrinsic paucity of operable slip systems in the hexagonal close-packed (HCP) lattice restricts their ductility and formability at ambient temperatures, creating a persistent trade-off between strength and plasticity [5-10]. Microstructural control through grain boundary (GB) engineering has therefore emerged as a key strategy for improving mechanical performance.[11-16] Achieving such control requires a tailored recrystallization process with regulated grain growth—and consequently well-defined grain size and texture [17-19], which ultimately depends on the mobility of lattice defects, particularly dislocations and grain boundaries[20-26].

Static and dynamic recrystallization (SRX/DRX) in Mg alloys proceed through a hierarchical evolution of crystalline defects. During the nucleation stage, high density of dislocations migrate and re-arrange into low energy configurations, typically forming low-angle grain boundaries (LAGBs). As recrystallization progresses, the LAGBs continue to migrate and absorb dislocations, eventually transition to high-angle grain boundary (HAGB), which governs the growth and eventual stabilization of recrystallized grains [27-29]. Micro-alloying is therefore widely employed to regulate recrystallization by manipulating defect mobility and thereby retarding GB migration [30-32]. However, because LAGBs and HAGBs dominate different stages of recrystallization, their respective responses to alloying-induced interactions may differ significantly, underscoring the importance of evaluating their mobilities independently.

Alloying additions influence boundary migration through two primary pathways: interactions between dissolved atoms and migrating interfaces, and pinning effects arising from second-phase particles [33-36]. Existing studies on Mg alloys have predominantly focused on the behavior and strengthening roles of precipitates [37-40], whereas the contribution of dissolved alloying elements—particularly their interactions with different types of grain boundaries—has received comparatively less attention. Because solute effects and precipitation often coexist during thermomechanical processing, experimentally isolating their respective influences on defect mobility remains challenging [41]. This issue is especially critical in recrystallization, where LAGBs and HAGBs govern different stages of microstructure evolution and may exhibit distinct sensitivities to alloying-induced interactions. These considerations highlight the need for approaches capable of independently evaluating how alloying additions in both solid-solution and precipitated states affect the mobility of LAGBs and HAGBs.

Experimental studies have nevertheless demonstrated that alloying additions can significantly influence recrystallization behavior in Mg alloys[41-45]. Segregating elements such as Ca, Gd, and various rare-earth additions have been shown to suppress recrystallization, stabilize substructures, and increase GB slip resistance. For instance, Wang et al.[42] demonstrated that Gd addition reduced recrystallized grain size and enhanced strength in Mg–4Zn–0.6Ca alloys; Zeng et al.[41] observed a significant increase in low-angle GB density in Ca-modified AZ31 alloys, confirming that Ca segregation hinders GB slip; Robson et al.[43] attributed the suppression of DRX by rare earth elements to GB segregation-induced slip resistance. Such findings clearly indicate the importance of alloying-induced GB interactions. However, conventional characterization techniques lack the resolution required to quantify atomic-scale drag forces or to decouple the contributions from different boundary types. As a result, the fundamental mechanisms by which alloying additions influence the migration of LAGBs and HAGBs in Mg alloys remain poorly understood.

Atomistic simulations offer a powerful means of overcoming these limitations. In face-centered cubic alloys such as Al–Mg and Al–Cu systems, molecular dynamics (MD) studies have successfully revealed segregation behavior, solute-induced reductions in excess free volume, and the associated drag forces exerted on migrating GBs[46-49]. Importantly, these studies have demonstrated that allowing solute segregation—rather than assuming an idealized random solid-solution state—is essential for correctly assessing the relative contributions of elastic and chemical pinning mechanisms. In contrast, analogous atomic-scale analyses in HCP Mg alloys remain scarce, despite their greater structural complexity and pronounced sensitivity to alloying additions. In particular, a systematic quantification of how alloying elements differentially affect the migration of LAGBs versus HAGBs is still lacking—representing a critical gap for establishing mechanistic links to the stage-dependent nature of recrystallization.

In this study, we employ atomic-scale simulations to quantitatively decouple and elucidate the retarding mechanisms of solute atoms on various crystalline defects in Mg. We select pure Mg as the reference matrix and investigate Al and Ca as model solutes due to their distinct strengthening roles: Al as a classical solid-solution strengthener [50, 51] and Ca as a potent segregator known for enhancing plasticity [52, 53]. By systematically evaluating the stress imposed by random and segregated solutes on $\langle 11\bar{2}0 \rangle$ edge dislocations, LAGBs, and characteristic HAGBs under both random and segregated solute configurations, this study aims to decouple elastic solute drag from chemically driven pinning mechanisms. Particular emphasis is placed on elucidating the distinct sensitivities of LAGBs and HAGBs to solute interactions, thereby establishing atomic-scale links between solute-controlled defect mobility and the stage-dependent evolution of recrystallization microstructures in Mg alloys.

## 2. Computational method

Molecular dynamics simulations were conducted using the Large-scale Atomic/Molecular Massively Parallel Simulator (LAMMPS) package [54], with visualization of results performed using the OVITO software [55]. We adopted the interatomic potential developed by Jang et al.[56]. This potential accurately reproduces critical properties of Mg-Al and Mg-Ca systems including cohesive energies and lattice constants—demonstrating excellent agreement with experimental measurements and first-principles calculations. Its reliability has been further validated across multiple studies [57-60].

To investigate the behavior of an individual dislocation, a rectangular single-edge-dislocation model measuring $200 \times 250 \times 67 \text{ Å}^3$ was constructed (Fig. 1a). The edge dislocation model was generated by joining and relaxing two grains with a one-atomic-layer height difference on the $(0001)$ plane, containing a single $\langle 11\bar{2}0 \rangle$ edge dislocation with the Burgers vector $\boldsymbol{b} = \langle 11\bar{2}0 \rangle$ on the $(11\bar{2}0)$ slip plane. Periodic boundary conditions were applied along the dislocation line direction (Z-axis) and the slip plane normal (X-axis), while free-surface boundary conditions were imposed along the Y-axis. To minimize boundary effects from the outer surfaces, a 7 Å-thick atomic layer on the Y-facing surfaces was fixed, allowing only atoms in the inner region to relax. The system was equilibrated for 0.01 ns under the NPT ensemble.

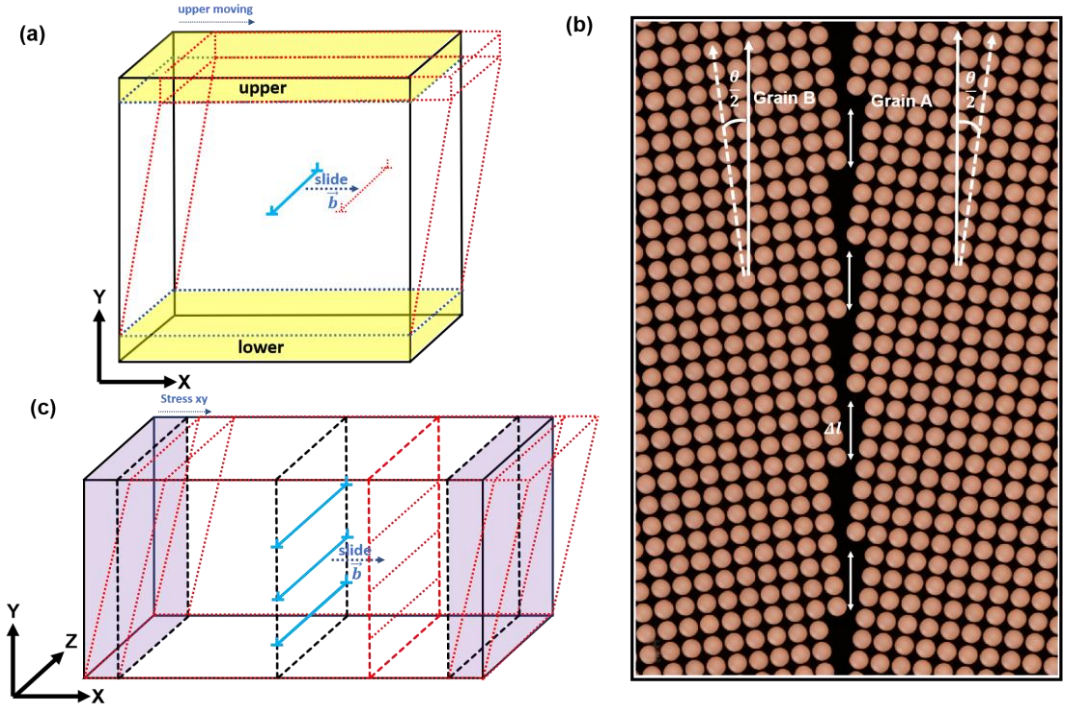

**Fig. 1.** (a) Configuration of the edge dislocation model under shear loading, where yellow regions denote the fixed atomic layers at the upper and lower boundaries, and red dashed lines indicate the post-loading positions of the simulation box and dislocation; (b) Schematic illustration of grain boundary formation via grain rotation and attachment; (c) Geometry of the grain boundary model under shear loading, with purple regions representing the fixed atomic layers at the left and right boundaries, and red dashed lines showing the positions of the simulation box and grain boundary after loading.

Following the single-dislocation modeling, grain boundary (GB) models were constructed following the method in Ref. [61] by rotating two grains about a common axis along [10$\bar{1}$0] to specific misorientation angles, followed by translational alignment, structural joining, and relaxation (Fig.1b). LAGBs were generated with misorientations ranging from 2° to 15°, while four representative HAGBs were constructed at misorientations of 32.24°, 63.28°, 101.88°, and 123.18°, corresponding to GB planes along (11$\bar{2}$1), (11$\bar{2}$2), (11$\bar{2}$4), and (11$\bar{2}$6). The orthogonal simulation box was defined with initial basis vectors X // [11$\bar{2}$0], Y // [0001], and Z // [10$\bar{1}$0], and dimensions of $800 \times 160 \times 66$ Å$^3$. Periodic boundary conditions were applied along the Y and Z directions (Fig. 1c), whereas a 5 Å thick atomic layer near the surface along X directions was fixed to apply shear strain, allowing all inner atoms to relax freely. All models were subsequently equilibrated in the NPT ensemble at 623 K (the recrystallization temperature of magnesium alloys) [62-64] and zero pressure for 0.1 ns to ensure it reached thermodynamic equilibrium and that internal stresses were fully released.

Following the initial relaxation, solute atoms were introduced into the pure Mg matrix via two distinct approaches: random substitution of Mg atoms to achieve a uniform solid solution, and selective segregation of solutes to dislocation cores or grain boundary regions using the semi-grand canonical Monte Carlo (SGCMC) method. To accurately model solute segregation behavior, a hybrid simulation scheme coupling molecular dynamics with SGCMC was employed. This approach enables the establishment of thermodynamic equilibrium without performing full kinetic simulations, thereby preventing the issue of reduced solid solution concentration due to solute segregation. In the SGCMC simulations, solute distribution was controlled by setting a chemical potential difference, with Mg as the reference ($\mu_0 = 0$) and individually adjusting the chemical potentials of Al and Ca. The relationship between solute concentration $C_{solute}$ and $\Delta\mu = \mu_{solute} - \mu_0$ was first established using a defect-free HCP-Mg model over a range of chemical potentials. By systematically exploring $\Delta\mu$ values from 0 to –5 eV, chemical potentials corresponding to approximately 3 $at.\%$ solute concentration was selected: $\mu_{Al} = -1.916\ eV$ and $\mu_{Ca} = -0.636\ eV$. In the hybrid simulations, MD runs at 623 K and SGCMC sampling were alternated in a 10:1 ratio, performing 20 SGCMC steps after every 200 MD steps.

Following the construction of solute-containing models, shear loading simulations were conducted separately for the single $\langle 11\bar{2}0 \rangle$ edge dislocation and GB models. For the single $\langle 11\bar{2}0 \rangle$ edge dislocation model (Fig. 1b), a uniform displacement was applied to the upper region, tilting the simulation box in the mobile atomic zone and generating shear stress that drives dislocation glide along the $(0001)$ basal slip plane. For the GB model (Fig. 1c), simple shear was imposed by tilting the simulation box within the XY plane, enabling the migration of grain boundaries and dislocations along the $(0001)$ basal plane. In both cases, a strain rate of $0.1\ ns^{-1}$ was applied.

To accurately evaluate the interactions between solute atoms and their influence on segregation behavior, we calculated the binding energies between two defects using the following expression:

$$E_b = E_A + E_B - E_{A+B} - E_{bulk}. \tag{1}$$

Here, $E_A$ and $E_B$ correspond to the total energies of the respective single-defect systems; $E_{A+B}$ represents the total energy of the system containing two defects A and B within the Mg matrix (solutes, grain boundaries, dislocations, etc.); $E_{bulk}$ is the total energy of the pristine Mg matrix.

Apart from MD simulations, we performed benchmark density function theory (DFT) calculations to confirm the accuracy of the interatomic potential using the Vienna Ab initio Simulation Package (VASP) [65, 66] with the projector augmented-wave (PAW) method [67]. The d and s electrons of all metallic elements were treated as valence electrons. The exchange-correlation functional was treated within the generalized gradient approximation (GGA) using the Perdew Burke Ernzerhof (PBE) parameterization. Three supercells containing 108 atomic sites (constructed as 3×3×3 replicas of the rectangular HCP cell with basis vectors X ∥ [11$\bar{2}$0], Y ∥ [10$\bar{1}$0], and Z ∥ [0001])were employed in the calculations. A plane-wave cutoff energy of 350 eV was used with convergence thresholds of $10^{-7}$ eV for energy and 0.01 eV/Å for atomic forces. A 4×3×3 k-point mesh generated by the Monkhorst-Pack scheme [68] was adopted with a smearing width of 0.1 eV. All atomic configurations were fully relaxed, and the cell shape and volume were optimized.

## 3. Results

### 3.1 Slip behavior of a single ⟨11$\bar{2}$0⟩ edge dislocation

We start by examining the fundamental behavior of dislocation slip in pure Mg, using it as the reference matrix to establish a baseline for solute effects. The slip behavior of ⟨11$\bar{2}$0⟩ edge dislocation in pure Mg was first characterized as a function of temperature. As illustrated in Fig. 2, the process proceeds through characteristic atomic shear slip accompanied by distinct system energy variations. The corresponding flow stress exhibits a pronounced temperature dependence, increasing from 0.0078 MPa at 0 K to 10.1 MPa at 300 K and 18.3 MPa at 623 K. This behavior reflects the influence of phonon drag on dislocation motion: at elevated temperatures, dislocation motion is hindered by phonon scattering, which imposes a very significant drag stress in comparison to the 0 K flow stress. As temperature decreases, this contribution to the resistance diminishes rapidly [69, 70]. Consequently, a higher applied stress is required to overcome the drag and sustain dislocation glide.

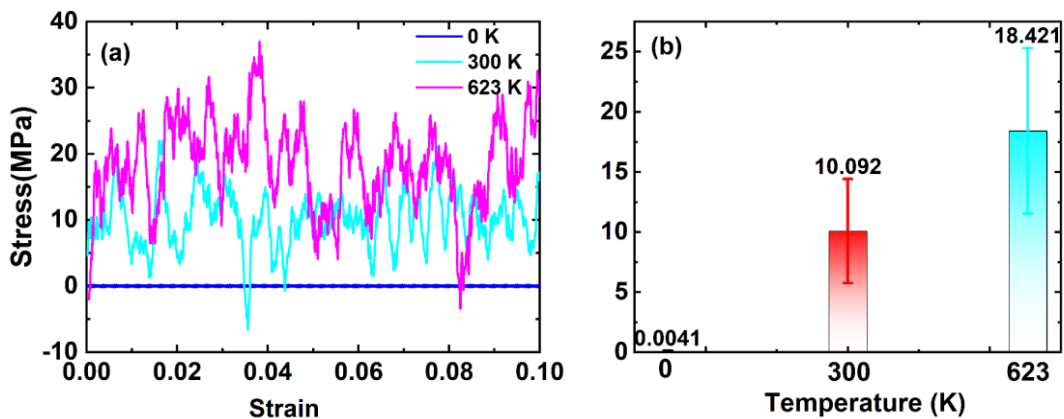

**Fig. 2.** (a) Stress-strain response of ⟨11$\bar{2}$0⟩ edge dislocation under shear deformation at 0 K, 300 K, and 623 K; (b) Average stress required for glide of ⟨11$\bar{2}$0⟩ edge dislocation at different temperatures (0 K, 300 K, 623 K), error bars denote standard deviation.

Having established the intrinsic behavior of dislocation slip in pure Mg, the focus shifts to how solutes perturb this process. To probe solute effects, we constructed models of an ⟨11$\bar{2}$0⟩ edge dislocation in binary Mg-Al and Mg-Ca system, considering both randomly distributed and GB-segregated solute configurations (obtained by MC simulation) at 3 $at.\%$ concentration. As shown in Fig. 3, the impact of solutes on ⟨11$\bar{2}$0⟩ edge dislocation slip is highly sensitive to their distribution state. In systems with randomly distributed solutes, both Al and Ca significantly increase the flow stress relative to pure Mg (18.3 MPa), with Al increasing it to 40 MPa, while Ca pushing it further to 50 MPa, underscoring a stronger impeding effect from Ca. This difference primarily stems from the pronounced lattice distortion (volumetric effect) induced by the larger atomic radius of Ca (Table 1), which will couple with the stress field induced by dislocation, resulting in a solute-dislocation attraction. This is further corroborated by the calculated binding energies of solute atoms within the dislocation core region. As demonstrated in Fig. 4, Ca atoms exhibit significantly higher binding energies to the dislocation core than Al, with a difference of 0.02–0.3 eV in the atomic layer from -6 to -1, stabilizing their pinning in the highly strained core region and thereby exerting stronger resistance to dislocation slipping.

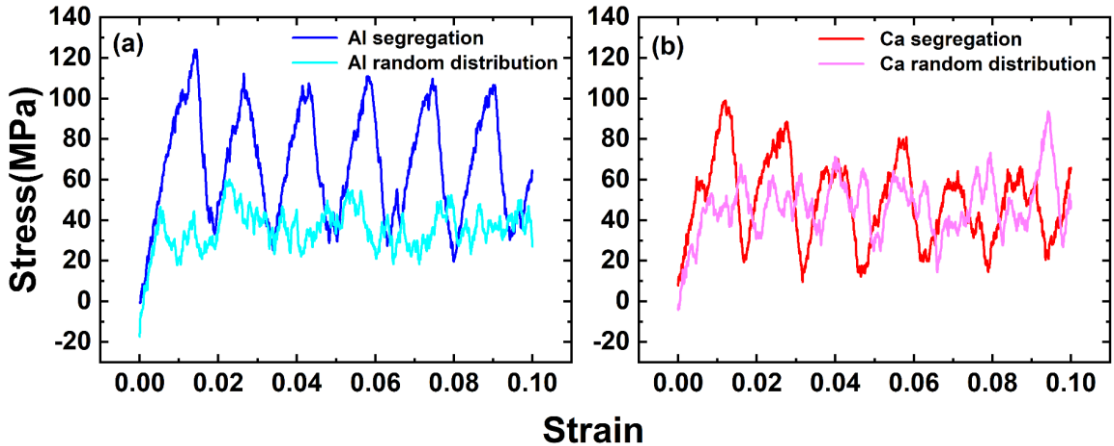

**Fig. 3.** Shear stress-strain curves of ⟨11$\bar{2}$0⟩ edge dislocation in Mg with 3 $at.\%$ of Al (a) and Ca (b) solutes, comparing cases with randomly distributed solutes and GB-segregated configurations.

**Table 1.** Binding energies of Al-Al and Ca-Ca atomic pairs at first and second nearest-neighbor sites from MD/DFT calculations, and volume changes induced by Al/Ca solutes in pure Mg lattice.

| Atom type | MD | | | DFT | | |
|---|---|---|---|---|---|---|
| | $E_b^{1nn}$ (eV) | $E_b^{2nn}$ (eV) | $\Delta V$ (Å³) | $E_b^{1nn}$ (eV) | $E_b^{2nn}$ (eV) | $\Delta V$ (Å³) |
| Al | -0.012 | -0.015 | -6.97 | -0.016 | -0.018 | -5.63 |
| Ca | -0.167 | -0.159 | 13.91 | -0.096 | -0.107 | 18.01 |

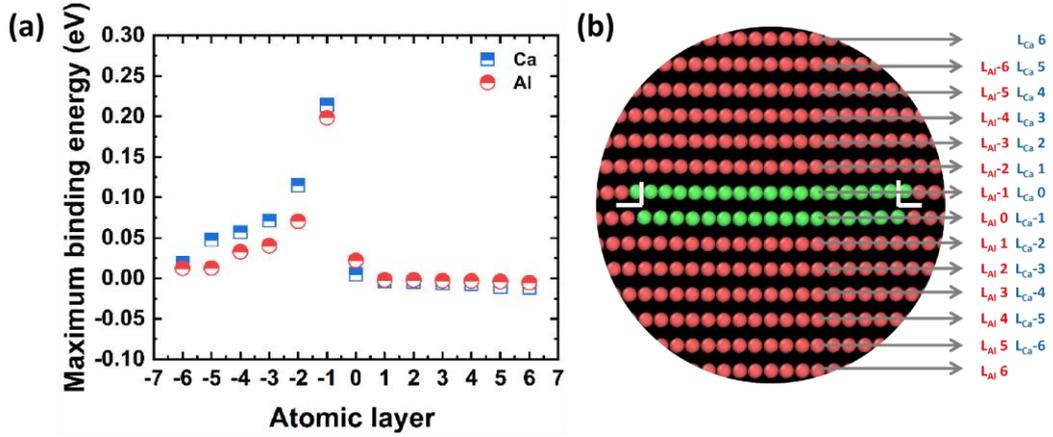

**Fig. 4.** (a)Binding energies of Al and Ca solutes to an $\langle 11\bar{2}0 \rangle$ edge dislocation in Mg. Only the maximum binding energy values at each atomic layer along the dislocation core are presented. Al atoms preferentially bind on the compressive side of the dislocation, whereas Ca atoms favor the tensile side. For direct comparison, the Al data are mirrored about the slip plane. (b) Schematic of local atomic configurations and layer definitions near the dislocation core. Atoms are colored by their local coordination: red for HCP and green for FCC.

It is worth noting that the random distribution of solutes, though widely adopted in previous studies for its simplicity, may represent an oversimplification of realistic alloying conditions, where solute segregation is frequently observed. Further analysis of the data in Fig. 3 reveals a contrasting influence of segregated solutes on dislocation motion between Al and Ca. The critical slip stress for the dislocation containing segregated Al reaches 120 MPa—substantially higher than the 100 MPa measured for its Ca-segregated counterpart. Moreover, the slip process in the presence of segregation exhibits a distinct periodic pinning–depinning behavior (Fig. 3), arising from repeated interactions between the migrating dislocation and solute clusters as the dislocation traverses the periodic boundaries. This oscillatory response is a hallmark of strong solute pinning effects.

The stronger impediment associated with Al segregation can be attributed to its markedly higher local concentration. As shown in Fig. 5, after equilibration at 623 K, approximately 800 Al atoms segregate to the $\langle 11\bar{2}0 \rangle$ edge dislocation core, a population fourfold greater than the ~200 Ca atoms observed under equivalent conditions. This divergence in segregation potency cannot be ascribed to volumetric effects alone; although Ca induces a more significant lattice expansion compared to the contraction of Al, its efficacy is ultimately limited by solute-solute interactions. Our combined MD and DFT calculations elucidate this critical distinction. The binding energy for Al-Al pairs is marginally repulsive, ranging from -0.01 to -0.02 eV, indicating negligible repulsion that permits dense segregation. In stark contrast, Ca-Ca interactions are strongly repulsive, with binding energies around -0.1 eV. This substantial repulsive force imposes a ceiling on the enrichment of Ca within the core region. Consequently, under segregated conditions, Al atoms achieve a stronger chemical pinning effect through their higher local concentration.

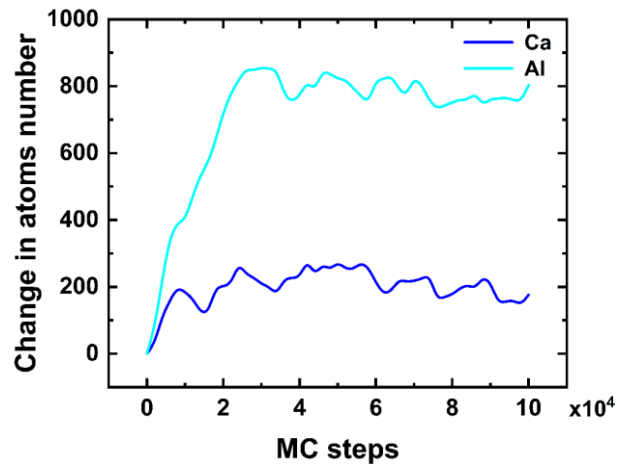

**Fig. 5.** Evolution of Al/Ca solute atom populations segregated at the $\langle 11\bar{2}0 \rangle$ edge dislocation during MC simulations.

**3.2 Slip behavior of LAGBs**

Building upon the elucidated slip mechanisms of the $\langle 11\bar{2}0 \rangle$ edge dislocation, we further investigated the slip behavior of LAGBs with misorientation angles ranging from 4° to 15°, which are composed of regularly spaced $\langle 11\bar{2}0 \rangle$ edge dislocations. As shown in Fig. 6, the grain boundary energy increases monotonically with misorientation angle and follows the Read-Shockley relation [71] with remarkable accuracy, confirming both the structural fidelity and physical validity of our atomistic model. This correlation demonstrates that the simulated LAGBs effectively capture the elastic interaction and contributions governing boundary energetics in Mg.

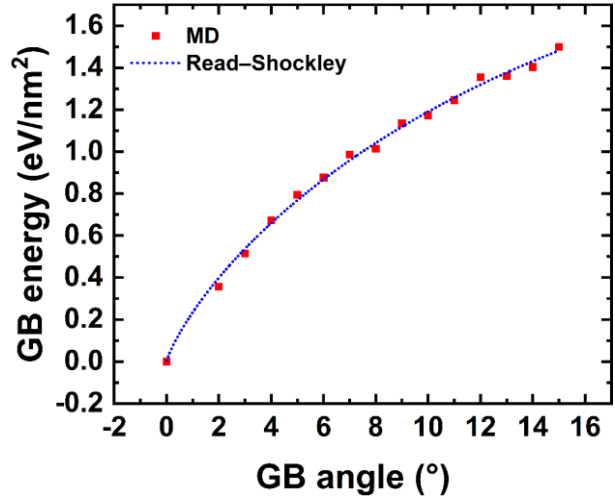

**Fig. 6.** Energy evolution of LAGBs with misorientation angles from 0° to 15°.

To extend this analysis to solute effects, the resistance to slip under different solute distribution states at 623 K was quantitatively characterized by applying a constant shear strain rate. As shown in Fig. 7, the slip behavior of LAGBs exhibits little dependence on misorientation angle in all systems—whether in pure Mg, or models with randomly distributed or segregated solutes. In pure Mg models, the stress required to initiate LAGB slip is relatively low, with the entire migration process occurring under stresses between 0 and 10 MPa. Compared to the solute-free case, the presence of randomly distributed Al or Ca solutes leads to a moderate increase in the slip stress. Notably, Ca exhibits a much stronger impeding effect than Al, consistent with the behavior observed for isolated $\langle11\bar{2}0\rangle$ edge dislocations. This enhanced resistance originates from the pronounced lattice distortion and associated elastic interaction introduced by the larger Ca solute atoms, as summarized in Table 1. The oversized Ca atoms interplay with the local strain field near the dislocation core, thereby increasing the energy barrier for dislocation motion. In contrast, Al with smaller volumetric misfits generates weaker elastic interactions and consequently exerts a milder strengthening effect. Taken together, these results reveal that the solute–dislocation interaction mechanisms established at the single-dislocation level are preserved in more complex defect configurations such as LAGBs. The magnitude of solute-induced strengthening thus scales with both the intrinsic misfit strain and the local solute concentration at dislocation cores, providing a microscopic basis for solute-tuned grain boundary strengthening in Mg alloys.

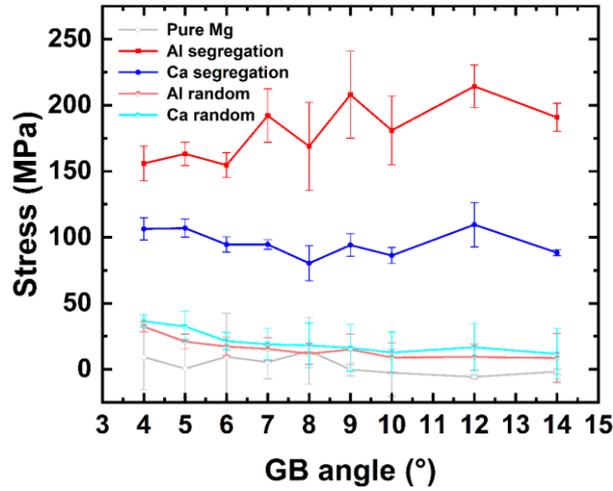

**Fig. 7.** Slip stresses of LAGBs (4°–14°) in pure Mg, and Mg with randomly distributed/segregated Al or Ca solutes.

Conversely, when solute atoms segregate to the LAGB, the slip resistance becomes primarily governed by the local solute concentration at the boundary rather than by lattice distortion alone. Under these conditions, a clear reversal in the solute hindrance effect is observed—the strengthening effect of Al surpasses that of Ca. This transition highlights the critical role of segregation thermodynamics in determining boundary mobility. As shown in Fig. 8, at 623 K, the equilibrium segregation concentration of Al in the grain boundary core region reaches approximately 4.5 $at.\%$, notably higher than the 3.5 $at.\%$ observed for Ca.

This contrasting segregation tendency stems from the intrinsic differences in solute–solute interaction energies (Table 1). The relatively weak Al–Al repulsion, reflected in binding energies ranging from −0.012 eV to −0.018 eV, allows Al atoms to accumulate densely along the boundary plane, thereby promoting enhanced chemical pinning of the LAGB. In contrast, Ca atoms exhibit much stronger mutual repulsion, with binding energies between −0.096 eV and −0.167 eV, which limits their attainable segregation level despite their larger size. As a result, although Ca induces greater lattice distortion in the Mg matrix, its ability to impede boundary motion diminishes under segregation-dominated conditions. The competition between chemical and volumetric effects thus explains the crossover behavior in solute strengthening between Al and Ca at the LAGB.

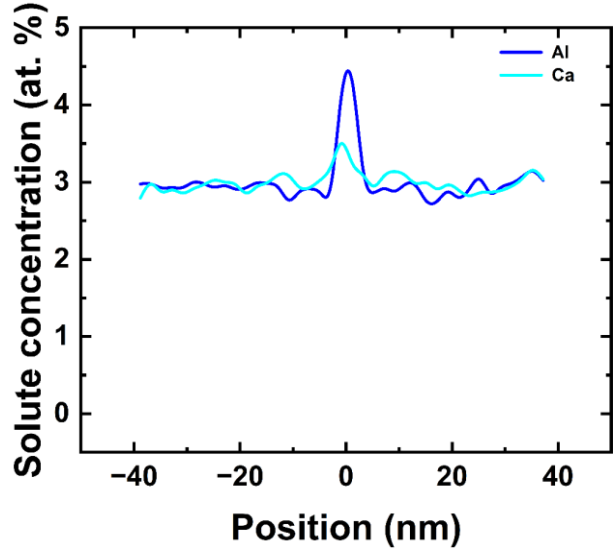

**Fig. 8.** Spatially resolved concentration profiles of Al/Ca solute atoms across LAGB regions. The grain boundary plane is centered at x=0.

The slip resistance of LAGBs follows a trend consistent with that observed for isolated $\langle 11\bar{2}0 \rangle$ edge dislocations, reinforcing the mechanistic continuity across scales. Specifically, (1) under random solute distributions, slip resistance is dominated by the volumetric effect, leading to a stronger hindrance from Ca than from Al; and (2) under solute segregation, resistance is primarily governed by the local solute concentration, reversing the trend to Al > Ca. This systematic correlation demonstrates that the slip behavior of LAGBs is fundamentally dictated by the collective glide of their constituent $\langle 11\bar{2}0 \rangle$ edge dislocation arrays, linking atomic-scale solute–dislocation interactions to mesoscopic grain boundary plasticity.

**3.2 Slip behavior of HAGBs**

Building upon the established slip mechanisms for the $\langle 11\bar{2}0 \rangle$ edge dislocation and LAGBs, we further extended our investigation to HAGBs to explore how increasing boundary misorientation influences solute–boundary interactions and slip behavior. Four representative low-energy HAGBs were selected for this purpose, characterized by misorientation angles and GB planes of 32.24° $(11\bar{2}1)$, 63.28° $(11\bar{2}2)$, 101.88° $(11\bar{2}4)$, and 123.18° $(11\bar{2}6)$ in $[10\bar{1}0]$ tilt axis. As show in Fig. 9, these boundaries corresponding to local energy minima in the energy–misorientation map, consistent with their structural characteristics as low-energy stable interfaces.

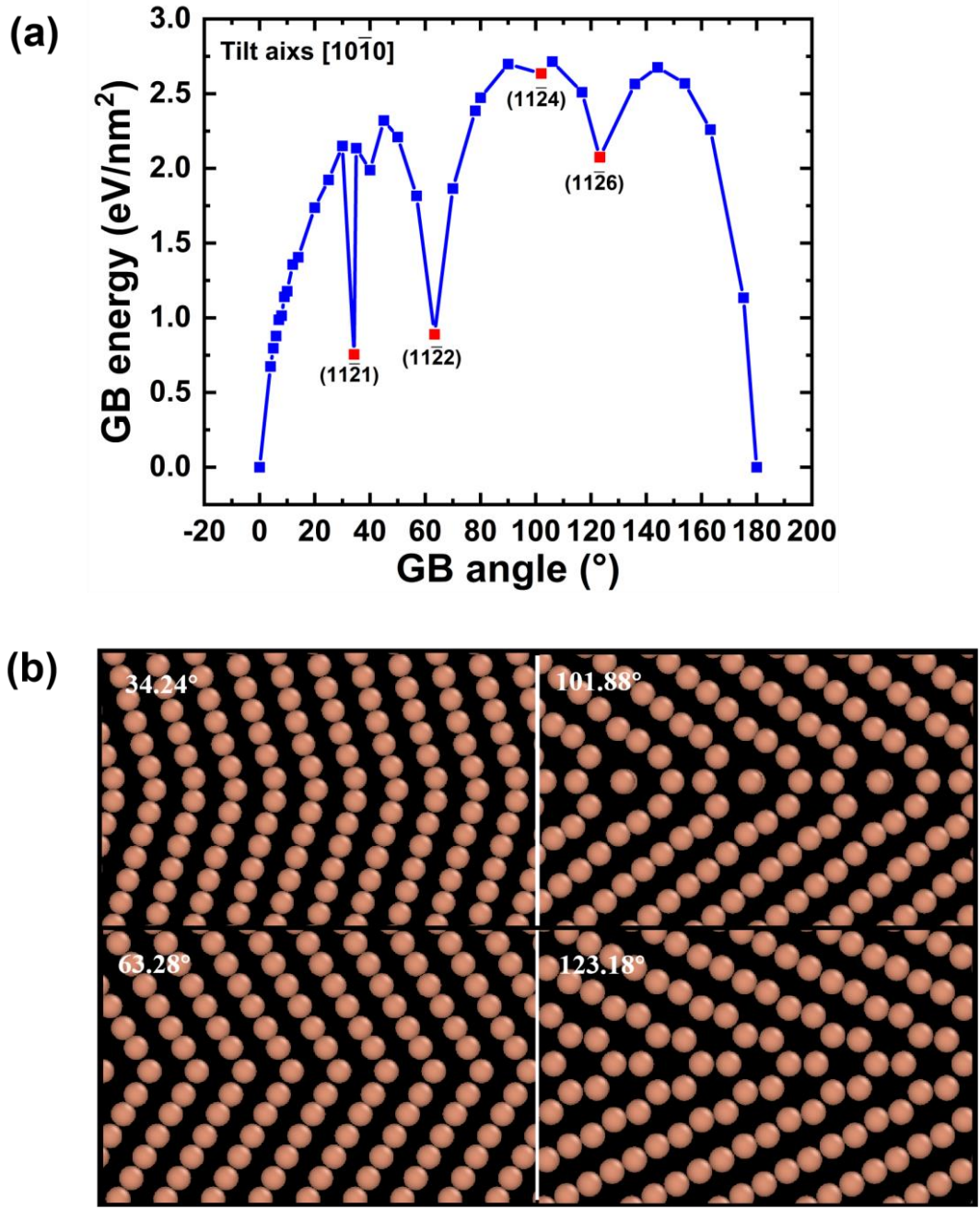

**Fig. 9.** (a) Evolution of GB energy as a function of misorientation angle in $[10\bar{1}0]$ tilt axis, ranging from LAGBs to HAGBs. Four representative HAGBs investigated in this work are highlighted by red symbols. (b) Corresponding atomic structures of the four HAGBs at characteristic misorientation angles.

For HAGBs containing randomly distributed solutes, the slip stress was slightly increased by both Al and Ca solutes for the 32.24°, 101.88°, and 123.18° boundaries, as demonstrated in Fig. 10. The impediment effect of Ca remains marginally stronger than that of Al, aligning with observations for LAGBs and arising from the pronounced lattice distortion induced by Ca (Table 1). In contrast, the 63.28° boundary exhibited a distinct response, where randomly distributed solutes slightly facilitated boundary slip.

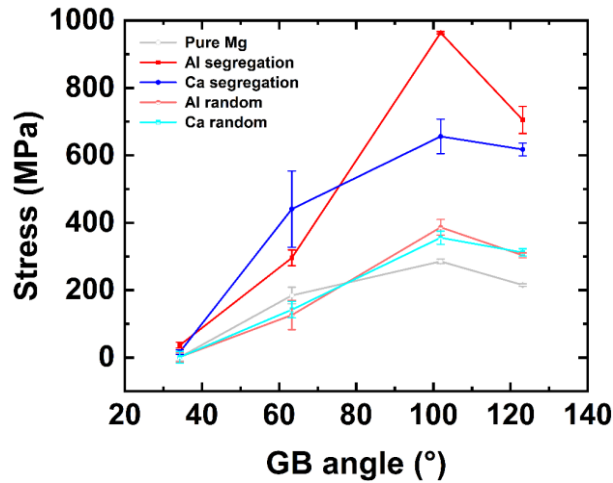

**Fig. 10.** Slip stresses for HAGBs in pure Mg and Mg-Al/Ca systems with random or segregated solutes at four misorientations: 32.24°, 63.28°, 101.88°, 123.18°.

Under solute segregation conditions, however, the slip response of HAGBs became more complex, reflecting a combined influence of volumetric effect and the local segregation concentration. For the 32.24° boundary, both Al and Ca exerted negligible effects on slip resistance due to intrinsically weak segregation tendencies. Analysis of binding energies in Fig. 10 shows values as low as 0.05 eV for Al and nearly zero for Ca, resulting in limited solute enrichment. Although the segregation concentration of Al slightly exceeded that of Ca, both remained close to the bulk concentration of 3 $at.\%$ showed in Fig. 12. As a result, both randomly distributed and segregated solutes increased the slip resistance by less than 10%, confirming the correlation between low segregation energy and weak solute drag.

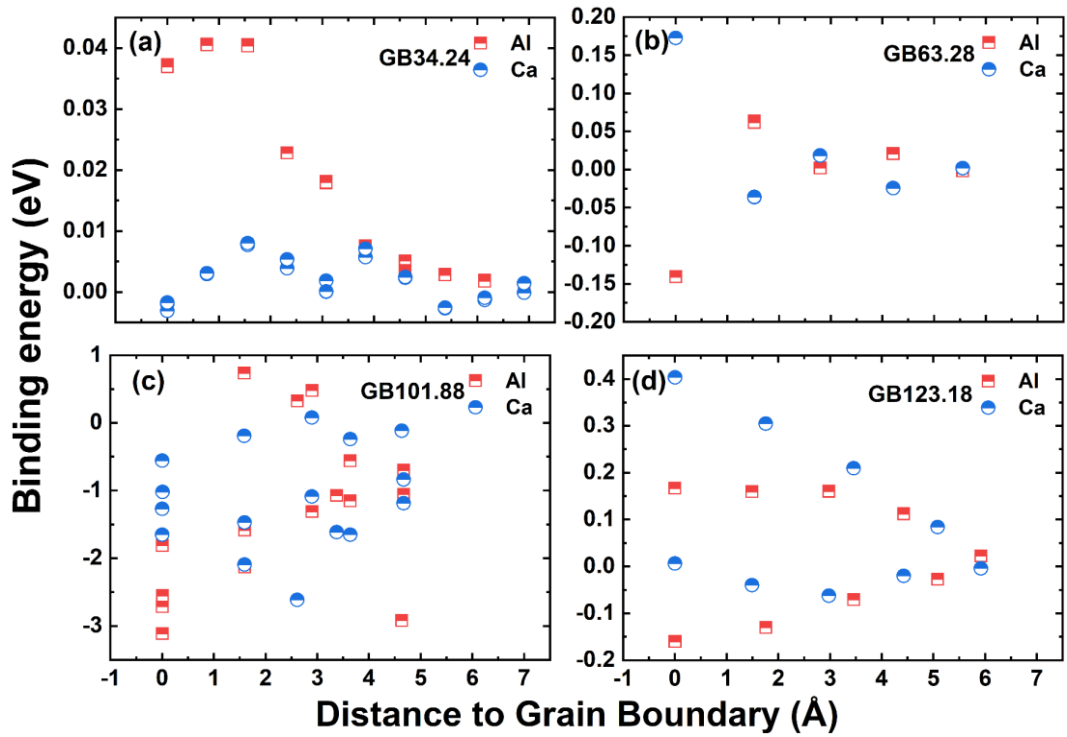

**Fig. 11.** Binding energy of Al and Ca atoms near HAGBs with misorientation angles of (a) 32.24°, (b) 63.28°, (c) 101.88°, and (d) 123.18°.

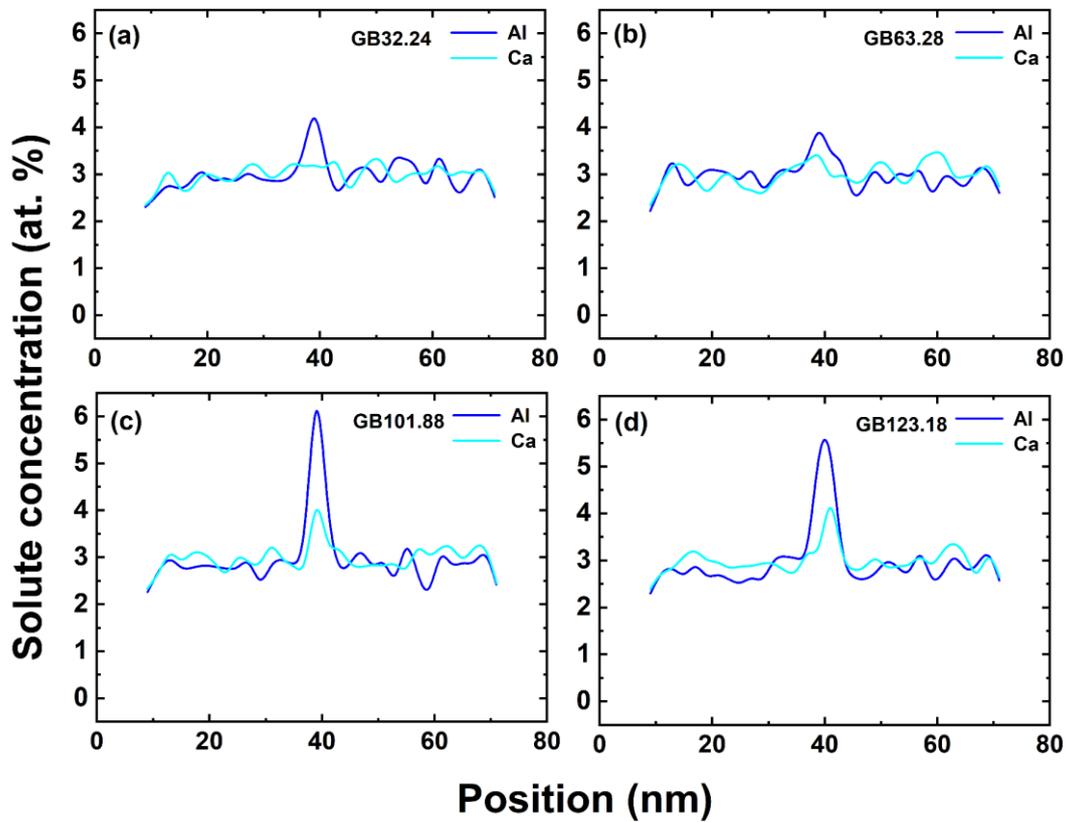

**Fig. 12.** Concentration profiles of Al and Ca atoms at HAGBs with misorientation angles of (a) 32.24°, (b) 63.28°, (c) 101.88°, and (d) 123.18°

A markedly different behavior emerged for the 63.28° boundary, which displayed a reversal of solute effects between random and segregated states. While randomly distributed solutes slightly facilitated boundary slip, segregated Ca induced a stronger impediment than Al. This apparent anomaly stems from the competition between segregation energy and achievable solute concentration, as illustrated in Figs. 11 and 12, Al exhibits negative segregation energy at the first nearest-neighbor site, indicative of repulsive segregation, whereas calcium possesses a positive binding energy of 0.15 eV. Despite this, both solutes reached comparable enrichment levels of approximately 3.5–4.0 $at.\%$. Strong repulsive interactions among Ca atoms, as documented in Table 1, limit its enrichment even with high segregation energy. Thus, Ca enhances pinning mainly through its volumetric effect, thereby compensating for its concentration disadvantage. In contrast, Al, although present at a slightly higher concentration, exhibits weaker pinning efficacy due to a less pronounced volumetric effect.

For the 101.88° boundary, the trend reverses, as segregated Al imposed a stronger impediment than Ca. Al enrichment reaches 6.0 $at.\%$, significantly higher than the 4.0 $at.\%$ enrichment of Ca. This enhanced segregation arises from higher segregation energy of Al, particularly at the second nearest-neighbor site, enabling the formation of a dense solute atmosphere that effectively pins the grain boundary. The chemical pinning effect from Al dominated the slip behavior, leading to an increase in slip stress that was 350–380 MPa higher than that induced by Ca. A similar but less pronounced trend is observed at the 123.18° boundary, where Al segregation also produced a stronger pinning effect than Ca. Although Ca has a higher overall segregation energy, strong repulsive interactions between Ca atoms limit its enrichment. As a result, Al reached a segregation concentration of 5.5 $at.\%$, considerably higher than the 4.0 $at.\%$ observed for Ca. The elevated local concentration of Al enhanced its chemical pinning effect, increasing the slip stress by 50–100 MPa compared to Ca.

Overall, the slip behavior across dislocation, LAGB, and HAGB hierarchies reveals a coherent solute–defect interaction framework in Mg. In all configurations, the dominant solute effect transitions from a volumetric mechanism under random distribution to a chemical pinning mechanism under segregation. The volumetric effect, governed by atomic size mismatch, renders Ca more effective at impeding defect motion in the absence of segregation. While the chemical pinning effect, controlled by segregation energy and solute–solute interactions, enables Al to dominate once concentrated at dislocation cores or grain boundaries. This mechanistic consistency across structural scales establishes a unified understanding of solute-controlled migration resistance in hcp Mg, providing atomistic insight for alloy design strategies that balance lattice distortion and segregation thermodynamics to optimize mechanical performance.

## 4. Discussion

The above results provide a mechanistic assessment of how Al and Ca solutes modulate the mobility of crystalline defects in Mg alloys across multiple length scales. A central finding is the markedly different sensitivity of LAGBs and HAGBs to solute-induced pinning. While the motion of LAGBs is only moderately affected by solute addition, HAGB migration exhibits a pronounced and highly solute-dependent resistance once segregation occurs. This contrast originates from the fundamentally different migration mechanisms governing these two boundary types and highlights the critical role of boundary character in determining solute drag effectiveness.

The origin of this differential sensitivity can be traced to a transition in the dominant solute–defect interaction mechanism, which is governed by the solute distribution state. Under homogeneous solid-solution conditions, solute drag is primarily elastic in nature, arising from atomic size mismatch between solute and matrix atoms. In this regime, Ca with larger atomic radius exerts a stronger resistance to the motion of dislocations and LAGBs. However, under realistic thermomechanical processing conditions relevant to recrystallization, solutes preferentially segregate to defect cores, shifting the controlling mechanism from elastic interactions to chemical pinning. Owing to their structurally disordered nature and larger excess free volume, partial HAGBs can accommodate substantially higher solute concentrations than LAGBs, thereby amplifying the effectiveness of chemical pinning at these interfaces.

Beyond solute–boundary affinity, solute–solute interactions emerge as a decisive factor controlling the ultimate pinning efficiency at HAGBs. Although an individual Ca atom induces a larger local lattice distortion, strong Ca–Ca repulsion severely limits the attainable segregation density at the boundary. In contrast, the weak interaction between Al atoms permits dense solute enrichment, enabling Al to generate a substantially stronger chemically driven pinning force once segregation occurs. This effect becomes particularly pronounced for HAGBs, where the high structural disorder facilitates solute accumulation and magnifies the impact of solute–solute interaction thermodynamics.

These computational insights establish a coherent mechanistic framework for interpreting key experimental observations. Both Nandy et al. [72] and Delis et al. [73] reported that introducing Al into Ca-containing Mg alloys, or increasing the Al content at fixed Ca concentration, leads to a pronounced reduction in the fully recrystallized grain size. According to our atomistic results, this behavior arises from the exceptional ability of Al to segregate densely at HAGBs owing to its weak solute–solute repulsion, which enables the formation of a concentrated solute atmosphere at the boundary. During the late stages of recrystallization, when grain growth is governed almost exclusively by HAGB migration, such chemically driven pinning by Al becomes particularly effective, thereby producing a strong and systematic grain refinement effect that scales with Al content.

Notably, Delis et al. [73] further observed that reducing the Ca content at a fixed Al concentration can also result in additional grain refinement, a trend that is mechanistically distinct from the Al-induced effects discussed above. Our simulations indicate that although Ca exhibits a strong atomic size misfit, its effectiveness in chemically pinning HAGBs is intrinsically limited by pronounced Ca–Ca repulsive interactions, which cap the attainable segregation concentration at the boundary. Moreover, experimental evidence indicates that increasing Ca content strongly promotes Al/Ca-containing precipitates [40, 74, 75], thereby depleting solute Al from the matrix and reducing its availability for segregation to HAGBs. Consequently, lowering the Ca concentration suppresses precipitation and releases more Al into solid solution, enabling enhanced Al segregation at HAGBs and strengthening chemical pinning. This competition between solute segregation and precipitation provides a coherent atomic-scale explanation for the counterintuitive grain refinement observed upon Ca reduction in Mg–Al–Ca alloys, suggesting the retardation mechanism of recrystallization is stage-dependent. This interpretation is further supported by the experimental study of Li et al. [40], who reported a significant increase in the fraction of LAGBs when the alloy composition was changed from Al 4.4 wt.%–Ca 1.1 wt.% to Al 4.0 wt.%–Ca 2.0 wt.%. The increased Ca content likely promotes Al/Ca precipitation and enhances elastic pinning of dislocation arrays, thereby stabilizing LAGBs during the early stages of recrystallization.

Taken together, the present results demonstrate that solute drag in hcp Mg alloys cannot be described by a single, universal mechanism. Instead, it arises from a hierarchy of interactions whose relative importance evolves with defect character and solute distribution (Fig. 13). The pronounced sensitivity of HAGBs to chemically driven pinning underscores their dominant role in controlling grain growth during the late stages of recrystallization. While solute segregation also induces chemical pinning of LAGBs, its effect is markedly weaker than that for HAGBs, suggesting early stage recrystallization likely dominated by precipitates upon co-alloying of Al and Ca. By explicitly decoupling these mechanisms at the atomic scale, this work provides a physically grounded framework for interpreting experimental recrystallization behavior and for guiding alloy design strategies that exploit targeted solute segregation to achieve stable, refined microstructures.

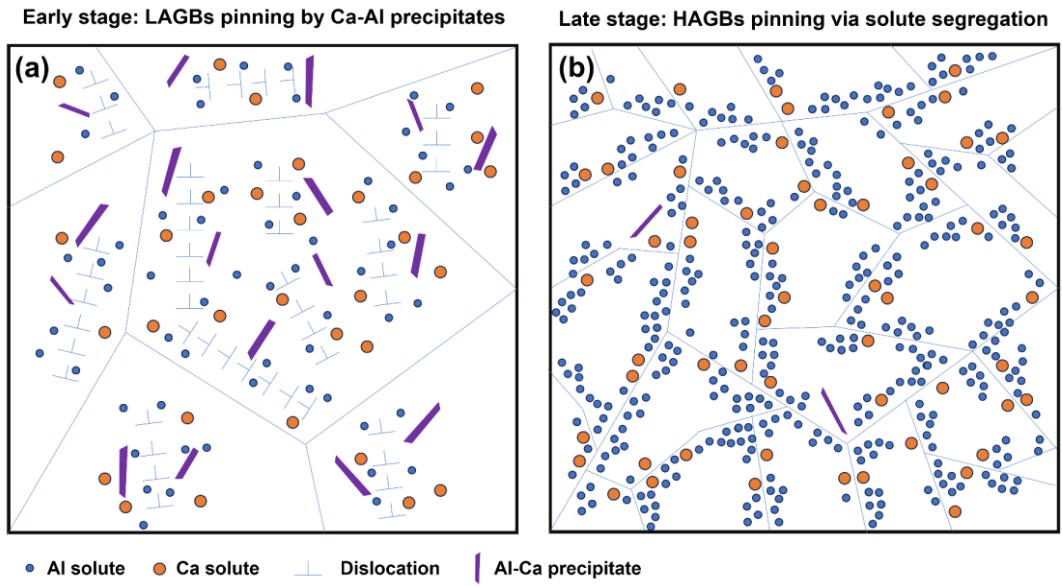

**Fig. 13.** Schematic diagram of the stage-dependent retardation mechanisms during recrystallization of Mg-Al-Ca alloys. Al-Ca precipitates (purple) primarily stabilize LAGBs in the early stages of recrystallization, whereas segregated Al (blue) and Ca (yellow) solutes exert enhanced chemical pinning on HAGBs during the later stages.

## 5. Conclusion

In this work, large-scale atomistic simulations were employed to systematically elucidate how solute distribution states regulate dislocation and grain boundary slip in Mg alloys. By explicitly comparing random solid-solution and segregated solute configurations, this study clarifies a long-standing ambiguity regarding the governing mechanisms of solute drag in Mg.

Under the commonly assumed but idealized random solid-solution condition, solute drag on basal $\langle11\bar{2}0\rangle$ edge dislocations and LAGBs is dominated by elastic interactions associated with atomic size mismatch, resulting in a stronger resistance from Ca than from Al. In contrast, in a more realistic condition considering solute segregations, the dominant mechanism shifts from elastic drag to chemically driven pinning, whose effectiveness is controlled by the attainable segregation concentration at defect cores. We demonstrate that weaker Al–Al repulsion enables substantially higher local enrichment of Al at dislocations and LAGBs compared with Ca, thereby reversing the relative pinning strengths of the two solutes.

More importantly, our results reveal a pronounced hierarchy in solute sensitivity between different grain boundary types. Although solute segregation modifies the retardation mechanism of LAGBs by introducing chemical pinning, its effect remains secondary. In contrast, HAGBs exhibit a much stronger response to solute-induced chemical pinning, indicating that solute effects are intrinsically amplified during the late stages of recrystallization and subsequent grain growth, where HAGB migration dominates microstructural evolution.

These findings establish a stage-dependent mechanistic framework for solute-controlled recrystallization in Mg alloys, offering atomic-scale guidance for alloy design strategies that exploit targeted solute segregation and composition optimization to achieve refined and thermally stable microstructures.

**Acknowledgement:**

This work was financially supported by National Natural Science Foundation of China (No.: 52401010; 12522517; 52471131), National Key R&D Program of China (No.: 2023YFB3710901), Hunan Provincial Natural Science Foundation of China (No.: 2025JJ40005; 2025JJ50219), Hunan Provincial Key Research and Development Program (No.: 2025QK3015), Natural Science Foundation of Jiangsu Province (No.: BK20240353), and the Fund of State Key Laboratory of Advanced Design and Manufacturing Technology for Vehicle (No.: 734215242). We acknowledge Hefei Advanced Computing Center for providing computing resources.

**Competing interests**

The authors declare no competing interests.

**Data availability**

The data generated and/or analyzed within the current study will be made available upon reasonable request to the authors.

**Author contributions**

**Zhishun Chen:** Methodology, Formal analysis, Investigation, Data Curation, Writing - Original Draft, Visualization. **Shudong He:** Writing - Review & Editing. **Shuai Zhang:** Writing - Review & Editing. **Xiaohan Bie:** Resources, Writing - review & editing. **Zhuoming Xie:** Resources, Writing - review & editing. **Tengfei Yang:** Writing - review & editing. **Wangyu Hu:** Resources, Supervision, Writing - review & editing. **Huiqiu Deng:** Writing - review & editing. **Shiwei Xu:** Writing - review & editing,

Supervision, Project administration. **Zhuoran Zeng:** Writing - review & editing, Supervision, Project administration. **Jie Hou:** Conceptualization, Methodology, Validation, Formal analysis, Resources, Data curation, Writing - Review & Editing, Supervision, Project administration, Funding acquisition.